\documentclass[12pt]{article} 

\usepackage{amsmath, amssymb, verbatim} 

\newcommand {\C } {\mathbb{C}} 

\newcommand {\p } {\mathbb{P}}

\newcommand {\h} {H^n(X,Y)} 

\newcommand {\Z} {\mathbb{Z}}

\begin{document} 

\title{\bf{Open string mirror maps from Picard- Fuchs equations on relative cohomology}} 

\author{Brian Forbes \\ \\ \it Department of Mathematics \\ \it University of California, Los Angeles \\ \it bforbes@math.ucla.edu, brian@math.sci.hokudai.ac.jp}

\date{July, 2003} 

\maketitle

\begin {abstract} 

A method for computing the open string mirror map and superpotential for noncompact Calabi-Yaus, following the physical computations of Lerche and Mayr, is presented. It is also shown that the obvious extension of these techniques to the compact case is not consistent. As an example, the local $\p^2$ case is worked out in 2 ways.

\end {abstract}

\section {Introduction.} 

Mirror symmetry has proven to be an effective tool in enumerative geometry. In Morrison's paper \cite{M}, it was shown how variations of physically motivated families of algebraic varieties led to predictions of the number of rational curves on the mirror varieties, through the study of Picard- Fuchs equations. The idea is that, by considering the $B$ model variation defined by a family of Calabi- Yau varieties, one can write down Picard- Fuchs equations and a basis of solutions for them. By a procedure of normalization on the PF solutions, it is possible to identify a canonical set of coordinates on the moduli space of the family of varieties, and these coordinates define the mirror map. One can then expand a certain function on the moduli space (called the prepotential $\cal F$) in terms of the canonical coordinates, and the coefficients of the expansion would give the predictions of rational curve numbers, up to a "multiple cover" formula, which was computed directly in 
\cite{LLY}. Some of these predictions were given explicit verification in terms of the mirror geometries in \cite{LLY}. 

More recently, there has been an interest in open string mirror symmetry and its enumerative consequences (\cite{LM}, \cite{LMW}, \cite{AV}, \cite{AKV}, \cite{GZ}, \cite{KL}). In terms of the $B$ model variation, or equivalently a family of algebraic varieties, the difference between the usual and the open string picture is that now one must vary a Calabi- Yau variety $X$ together with a holomorphic curve $B$ inside the variety. The new enumerative invariants, which still lack a definition (but see \cite{KL} for progress on this), are interpreted not as the number of rational curves in the mirror geometry, but rather as the number of Riemann surfaces with boundary whose boundary lies on a special Lagrangian submanifold of the mirror geometry.\cite{GZ} provides some mathematical exposition toward using the approach of \cite{LLY} on the open string problem. 

The focus of this paper is to use $B$ model computations to make predictions on the number of (conjectural) open Gromov- Witten invariants, in the spirit of Morrison's original paper\cite{M}. These ideas have already appeared in a physical setting, in the papers of Lerche, Mayr and Warner (\cite{LM}, \cite{LMW}). As in the paper \cite{M}, a $B$ model variation is used, which allows for the derivation of an extended set of Picard- Fuchs operators whose solutions give the open string mirror map. Also, the superpotential $W$ (which is analogous to the closed string prepotential $\cal F$) has been identified in \cite{LM}. This $W$, together with the open string mirror map and a new multiple cover formula \cite{KL}, allows for predictions of open Gromov- Witten invariants. Along the way, we will attempt to generalize the arguments in \cite{LMW} to the compact setting. As we will see, such a generalization is inconsistent; hence, there is some further subtlety at work in the compact open 
string case. 

Section 2 gives some mathematical background, which sets the stage for later computations. Section three explains the relevant spaces and the motivation for considering them; Section 4 identifies the appropriate relative cohomology class and GKZ operators. Section 5 carefully derives the PF operators and solutions, and section 6 contains the specific example $K_{\p^2}.$ Since most physical calculations deal with the noncompact case, this is presented explicitly; however, from a computational point of view the compact case is not vastly different, so compact results are presented throughout. 

\bigskip 

$\bf{Acknowledgements.}$ 

I would like to thank Wolfgang Lerche for patiently answering my questions about the superpotential computation. I also thank David Morrison and Matt Kerr for helpful conversations. Finally, thanks to my advisor, Kefeng Liu, for suggesting this direction, and to David Gieseker for support during the research. 

\bigskip

\section{Mixed Hodge structure and maximally unipotent points.}

\subsection{Mixed Hodge Structures.} 

Let $X$ be a smooth $n$ dimensional Kahler manifold, and $Y$ a codimension $1$ complex submanifold of $X$. In order to avoid primitive cohomology complications, choose $n$ to be odd. Here, all cohomology groups will have coefficients in $\C$, and all homology groups will have $\Z$ coefficients unless otherwise noted. $X$ and $Y$ carry Hodge structures of weight $n$ and $n-1$ respectively, described by the Hodge decompositions $$H^n(X) \cong \oplus_{p+q=n} H^{p,q}(X), \; H^{n-1}(Y) \cong \oplus_{p+q=n-1} H^{p,q}(Y)$$ together with the integer lattices $H^n(X,\Z)$, $H^{n-1}(Y,\Z)$. As usual, $H^{p,q}(X) = \overline{H^{q,p}(X)}$ , and the same for $Y$. Equivalently, one can talk about the Hodge filtrations 

$$F^{\bullet}(X) = \{F^p = \oplus_{a \geq p}H^{a,n-a}(X) \}_{p=0}^n$$ 
and the corresponding $F^{\bullet}(Y)$ , together with the above integer lattices.

The first goal is to define a mixed Hodge structure on the relative cohomology group $H^n(X,Y)$. The reason for this is the same as the usual setting; the mixed Hodge structure will tell us how to take derivatives of relative forms in $H^n(X,Y)$, and hence help identify the appropriate Picard- Fuchs operators. 

A mixed Hodge structure is defined by $(i)$ a complex vector space $V_{\C} = V_{\Z} \otimes {\C},$ where $V_{\Z}$ is a $\Z$- module, $(ii)$ an increasing weight filtration $\{W_k \}_{k \in I}$ of $V_{\Z} \otimes \mathbb{Q}$ , and $(iii)$ a decreasing Hodge filtration $\{F^p\}_{p \in J} $ of $V_{\C}$ for $I, J \subset \Z$ such that $\{(F^p \cap W_k) / (F^p \cap W_{k-1})\}_{p \in J}$ is a Hodge structure of weight $k$ on $W_k / W_{k-1}$. 

Later in the paper, we will take advantage of the isomorphisms $H^k(X,Y) \cong H^k_c(X-Y)$, that is, $k$ forms with compact support on $X-Y$. Also, from here forward we will be considering the simplified situation in which the long exact sequence on relative cohomology splits, so that 

\begin{equation} 
\label{eqno1} 
0 \longrightarrow H^{n-1}(Y) \longrightarrow \h \longrightarrow H^n(X) \longrightarrow 0 
\end{equation} 
is exact.

The mixed Hodge structure for this case can be found in \cite{GMV}. Simply take the Hodge filtration to be 

$$F^{\bullet} = \{F^p = \oplus_{a \geq p}H^{a,n-a}(X,Y) \}_{p=0}^n$$ 
together with the weight filtration $W_{n-1}, W_{n-2}$ given as 

$$W_{n-1} = \h, \ W_{n-2} = H^{n-1}(Y).$$ 
Then $$(F^p \cap W_{n-1}) / (F^p \cap W_{n-2}) = F^p(Y), \ p = 1 \dots n-1 $$ 
where the last equality comes from the short exact sequence (1). This is clearly a Hodge filtration on $H^{n-1}(Y)$, so we have a mixed Hodge structure on $\h$.

\subsection {Variation of Mixed Hodge Structure.}

Let $S$ be the punctured 2- disc, $S = (\Delta^*)^2$, with coordinates $z = (z_1, z_2)$. Let $\pi : \chi \longrightarrow S$ be a family pairs of smooth varieties defined by $\pi^{-1}(z) = (X_{z_1}, Y_{z_1, z_2})$, where $X_{z_1}$ is $n$ dimensional Kahler and $Y_{z_1, z_2}$ is a codimension 1 complex submanifold of $X_{z_1}$. For all $z \in S$, define $\h_z = H^n(X_{z_1}, Y_{z_1, z_2})$, and similarly for the cohomology groups on $X$ and $Y$. As in the previous section, assume the sequence 

$$ 0 \longrightarrow H^{n-1}(Y)_z \longrightarrow \h_z \longrightarrow H^n(X)_z \longrightarrow 0 $$ to be exact for each $z$. The mixed Hodge structure on each group $H^n(X,Y)_z$ is defined pointwise by the Hodge and weight filtrations given above. 

Next, we recall the definition of a variation of mixed Hodge structure (VMHS) \cite{SZ}: 

\bigskip 

\bf{Definition.} A \it variation of mixed Hodge structure \rm over a complex manifold $S$ consists of: 

(a) A local system $V_{\C}= V_{\Z} \otimes {\C}$ of $\C$ vector spaces on $S$ 

(b) An increasing filtration $W = \{W_k \}$ of $ V_{\mathbb{Q}} = V_{\Z} \otimes \mathbb{Q}$ by local subsystems 

(c) A decreasing filtration ${\cal F}^{\bullet} = \{{\cal F}^p \}$ of holomorphic subbundles of ${\cal F}^0 = V_{\C} \otimes {\cal O}_S$ 

(d) With respect to the canonical flat connection $\nabla$ defined on ${\cal F}^0$, $\nabla {\cal F}^p \subseteq {\cal F}^{p-1} \otimes \Omega_S^1$ 

(e) If ${\cal W}_k = W_k \otimes {\cal O}_S$, then $\{ ({\cal F}^p \cap {\cal W}_k) / ({\cal F}^p \cap {\cal W}_{k-1}) \}_p$ defines a variation of Hodge structure of weight $k$ on ${\cal W}_k / {\cal W}_{k-1}$. 

\bigskip 

A variation of Hodge structure is defined by the same information, less items (b) and (e). A discussion can be found in \cite{CK}. 

By setting $z_1 =c$ (resp.$z_2 =c$) for some constant $c$, it is known that the cohomology groups $H^{n-1}(Y)_{z_2}$ (resp.$H^n(X)_{z_1}$) vary over the base $S \mid_{z_1 = c}$($S \mid_{z_2 = c}$) in a locally constant fashion (\cite{S}). The splitting of the long exact relative cohomology sequence then gives that the variation of $\h_z$ over $S$ is also locally constant; hence $R^n \pi_* \C = V_{\C}$ defines a local system on $S$. It is then known from the general theory (\cite{SZ}) that the above family $\pi : \chi \longrightarrow S$ gives rise to a VMHS. For us, the relevant details of this are that for $z \in S$, the fiber of ${\cal F}^0 = V_{\C} \otimes {\cal O}_S$ is $\h_z$, and also that the fiber of ${\cal F}^p$ is $F^p_z$, the $p$th element of the Hodge filtration on $\h_z$.

\subsection { Maximally unipotent boundary points.}

The local system $V_{\C}$ is equivalent to a monodromy representation 

\begin{equation} 
\label{eqno2} 
\rho : \pi_1 (S, z_0) \longrightarrow GL(\h_{z_0}) 
\end{equation} 

In addition to the conditions imposed on $\pi : \chi \longrightarrow S$ above, we want to add that the representation (2) is unipotent. This is the definition of a unipotent VMHS. 

We would also like the VMHS to be \it admissible \rm in the sense of \cite{SZ}, which means, among other things, that the bundle ${\cal F}^0$ over $S$ extends to a bundle $\overline{{\cal F}^0}$ on $\overline{S} = \Delta^2$. Before this can be further explained, we need to recall some facts about relative weight filtrations (see \cite{SZ}).

So, let $V$ be a complex vector space with an increasing filtration $W$, and fix $k \in \Z$. Then the pair $(V, W)$ is a filtered vector space by definition. If also $N$ is a nilpotent endomorphism of $V$ which preserves $W$- that is, $NW_k \subset W_k$ for all $k$- then $N$ defines a linear map $$Gr_k N : Gr^W_k V \longrightarrow Gr^W_k V.$$ For any filtration $W$ of $V$, define, for any $n \in \Z$, a shifted filtration on $V$ by $W[n]_k = W_{k+n}.$ 

It is well known that a nilpotent endomorphism $N$ on $V$ determines a unique weight filtration $L$ on $V$; with the integer $k$ of the filtered vector space in hand, it is conventional to set $M(N) = L[-k]$. With these notations, one can make the following

\bigskip 

\textbf{Definition.}\textmd Let $(V,W)$ and $N$ be as above. A \itshape weight filtration of N relative to W \upshape is a filtration $M(N,W) = M$ of $V$ satisfying 

(1) $NM_k \subset M_{k-2},$ and 

(2) $MGr_k^W V = M(Gr_k N)$ 

for all k. 

\bigskip 

This comes with the useful 

\bigskip 

$\bf{Proposition.}$ There is at most one filtration $M$ of $V$ satisfying the conditions of the above definition. 

\bigskip 

For the monodromy representation (2) above, choose a basis $\{ T_j \}$ of Im$(\rho)$ and set $N = \Sigma_j a_j$ log $(T_j)$, with $a_j > 0$. Then $N$ is a nilpotent matrix. Let $W$ be the increasing weight filtration from the VMHS associated to the family $\pi : \chi \longrightarrow S$, and let $M = M(N,W)$. 

Since the VMHS associated to $\pi : \chi \longrightarrow S$ is geometric, $M$ is known to exist. Also, as mentioned earlier, in this case there is a canonical extention $\overline{{\cal F}^0}$ of ${\cal F}^0$ defined on $\overline{S}$. If $F_{lim}^{\bullet}$ denotes the limiting behavior of $\overline{{\cal F}^0}$ at $0$, then $(F_{lim}^{\bullet}, M)$ defines a mixed Hodge structure on the central fiber $\overline{\chi}(0)$. 

With these formalities out of the way, one is free to make the definition: 

\bigskip 

\textbf{Definition (Maximally unipotent boundary point for open strings).}\textmd A point $p \in \overline{S} - S$ is a \itshape maximally unipotent boundary point \upshape if the following are satisfied: 

(1) The representation $\rho : \pi_1 (S, z_0) \longrightarrow GL(\h_{z_0})$ for $z_0 \in S$ is unipotent. 

(2) If $N = \Sigma_j a_j$log$(T_j)$, where $\{T_j\}$ is a basis of Im$(\rho)$ and $a_j > 0$ for all $j$, and if $M$ is the weight filtration of $N$ relative to $W$, then dim$M_0 =$ dim $M_1 = 1$ and dim$M_2 = 3$. 

(3) Let $g_0$ span $M_0$ and extend this to a basis $\{g_0, g_1, g_2\}$ of $M_2$. If a matrix $(m_{ij})$ is defined by log$(T_i)(g_j) = m_{ij}g_0$, then $(m_{ij})$ is invertible. 

\bigskip 

Note that this is simply the usual definition of a maximally unipotent boundary point for $2$ dimensional moduli \cite{CK}, except that the monodromy weight filtration defined by $N$ has now been replaced by the filtration of $N$ relative to $W$.

\section {Simplifications for the toric case.}

\subsection{Compact Spaces.}

Now, we specialize to the case where the pair $(X,Y)$ are given as a Calabi- Yau hypersurface $X$ in a toric variety, together with a hypersuface $Y$ of $X$. In the interest of deriving the Picard- Fuchs equations in the simplest possible manner, one can enlarge the moduli space of the manifolds considered and take an appropriate quotient at the end; this is the strategy followed here. So, let $\Delta$ be a reflexive polytope of dimension $n+1$, and say $\p_{\Delta}$ is the toric variety defined by this polytope. If $N$ = Hom$(\Z^{n+1},\Z)$ and $T_N = N \otimes \C \subset \p_{\Delta},$ a choice of basis for $\Z^{n+1}$ selects coordinates $y_i$ on $T_N.$ With respect to this chosen basis, set $\{m_i\} = \Delta \cap {\Z}^{n+1}.$ 

Sections of the anticanonical bundle of $\p_{\Delta}$ can be described by Laurent polynomials 

$$f(a) = \sum_{i \ge 0} a_i y^{m_i},$$ 
where $y^{m_i} = {y_0}^{m_{i,0}}\cdot \cdot \cdot {y_n}^{m_{i,n}}$, the $y_i$ are as above and the $a_i$ are ${\C}^*$ variables. Then a family of Calabi- Yau toric hypersurfaces is given by $X_a$, which are defined as the zero sets of the polynomials $f(a)$. According to Batyrev \cite{B}, the mirror $\hat{X}$ of $X$ can be constructed from the dual polytope: if $\hat{\Delta}$ is the dual of $\Delta$, then by associating a Laurent polynomial to $\hat{\Delta}$ as above, we get the mirror family $\hat{X_a}.$ 

For reasons motivated by physics, the hypersurface $Y$ in $X$ will be defined in all cases to be 

$$Y_{a,b} = X_a \cap \{g(b) = b_0 y_0 + b_1 y_1 = 0\}.$$ 
Again, $b_i \in {\C}^*$ for $i= 0, 1.$ While this seems unmotivated at first, the next section on noncompact examples will clarify this choice. 

It should be noted that this is not the full generality presented on \cite{LMW}. There is in fact an integral ambiguity in the open string Gromov- Witten invariants, as discussed in \cite{AKV}. The above choice of $Y$ reflects 'ambiguity zero' in physics terminology. To include the ambiguity, fix some $N \in \Z.$ Then one would instead consider $Y$ to be given as 

$$Y_{a,b}^N = X_a \cap \{g_N(b) = b_0 y_0 + b_1 y_1^{N+1} y_2^{-N} = 0 \}.$$ 
Then the integer $N$ gives a discrete parameterization of the set of open Gromov- Witten invariants. For the rest of the paper, we will take $N = 0.$

To find coordinates $z_i$ such that the maximally unipotent boundary point is given by $z_i = 0$ for all $i$, the usual procedure for the determination of such coordinates in the closed string case is recalled here. In this context, 'closed string' refers to considering the moduli of $X_a$ only. One finds a set 

$$\Lambda = \{l \in {\Z}^{n+1}: \sum_{i=0}^n l_i m_i = 0, \ \ \ \sum_{i=0}^n l_{i} = 0 \}$$ 
of relations among the coefficients $\{m_i\}$ of $f(a)$. Here, there will be only one element of $\Lambda.$ Since $\Delta$ is reflexive, there is a Batyrev mirror of $X_a$, which gives a definite choice of basis vector for $\Lambda.$ Picking a basis vector $l^1$, there is a canonically defined coordinate on moduli space given as $z_1 = \prod_{k=0}^n {a_k}^{l_k^1}$, and this is conjectured to be the coordinate near the maximally unipotent point $z_1 = 0$. 

Next, we move on to the open string case, where the full moduli of $(X_a, Y_{a,b})$ is considered. One expects that there is a new set of relations $\widehat{\Lambda}$ which also take into account the moduli coordinates $b_i$, so that $\widehat{l} \in \widehat{\Lambda}$ are in ${\Z}^{n+3}$. The precise definition of this extended set $\widehat{\Lambda}$ will be made clear in subsequent sections. Once obtained, however, such $\widehat{l}$ will determine moduli coordinates just as in the closed string case, by the rule $z = \prod_{k=0}^n {a_k}^{{\widehat{l}}_k} {b_0}^{{\widehat{l}}_{n+1}} {b_1}^{{\widehat{l}}_{n+2}}.$

\subsection{Noncompact spaces.} 

Since the bulk of physical results concerning open Gromov- Witten invariants deal with noncompact Calabi- Yaus, we will need a formulation in terms of such spaces. Rather than directly write down the noncompact $\tilde X_a,$ we will first consider the mirror $A$ model spaces and apply the constructions of local mirror symmetry (\cite{CKYZ}, \cite{AKV}) to find the $B$ model hypersurfaces of interest. Along the way, we will gain a better understanding of the defining equation for $Y_{a,b}$ given above, which will allow for flexibility in adapting the presentation here to other geometries.

Following \cite{AKV}, the $A$ model Calabi- Yaus are given as symplectic quotients: 

\begin{equation} 
\label{noncptA} 
\tilde W_r = \{(x_0, x_1, x_2, x_3) \in (\C^4 - Z) | \sum_{i=0}^3 l_i |x_i|^2=r\} / S^1, 
\end{equation} 
where the $S^1$ action is 

$$ 
S^1 : x_i \longrightarrow e^{\sqrt{-1}l_i\theta}x_i 
$$ 
for $i=0\dots3$ and $r\in{\mathbb{R}}^+$. Note also that we need $\sum_{i=0}^3 l_i=0$ so that $c_1(\tilde W_r) = 0.$ Finally, $Z$ is the singular set implied by the relation $\sum_{i=0}^3 l_i |x_i|^2=r.$ That is, at least one $l_i =0,$ so suppose for example $l_0 <0$, and $l_i>0$ for $i\ne0$. Then $Z=\{x_1=x_2=x_3=0\},$ from the condition $r\in{\mathbb{R}}^+.$ 

Notice also that $r$ is the single Kahler parameter on this space, which is most easily seen through a concrete example. Set $l=(1,1,-1,-1).$ Then the symplectic quotient $\tilde W_r$ corresponding to this $l$ is known to be isomorphic to the total space of $\cal{O}$$(-1)\oplus\cal{O}$$(-1)\rightarrow\p^1.$ We have $Z=\{x_0=x_1=0\},$ and by setting $x_2=x_3=0$, we find 

$$ 
\tilde W_r \cap \{x_2=x_3=0\} = \{(x_0, x_1) \in \C^2 : |x_0|^2+|x_1|^2=r\} / S^1. 
$$ 

Since $\{(x_0, x_1) \in \C^2 : |x_0|^2+|x_1|^2=r\} \simeq S^3,$ we can identify $\{(x_0, x_1) \in \C^2 : |x_0|^2+|x_1|^2=r\} / S^1$ with $S^2 \simeq\p^1$ via the Hopf fibration. Then $r$ determines the size of the $\p^1$ in 
$\cal{O}$$(-1)\oplus\cal{O}$$(-1)\rightarrow\p^1,$ and this is the Kahler modulus on this space, as claimed. 

Recall that the usual definition of a Calabi- Yau manifold is a compact Kahler manifold with vanishing first Chern class. Clearly, the spaces $\tilde W_r$ are noncompact, since the coordinates corresponding to negative values of the vector $l$ are noncompact. This follows from the obvious generalization of the $\cal{O}$$(-1)\oplus\cal{O}$$(-1)\rightarrow\p^1$ example. Thus we relax the definition of Calabi- Yau, as is standard in physics literature, to say that any manifold which is Kahler and has vanishing first Chern class is Calabi- Yau. 

It is trivial to include extra Kahler parameters in the noncompact A model examples: 

$$ 
\tilde W_{r_1,\dots,r_k} = \{(x_0,\dots, x_{k+2}) \in (\C^{k+3} - Z) | \sum_{i=0}^{k+2} l_i^j |x_i|^2=r_j, j=1\dots k\} / (S^1)^k, 
$$ 
where now $Z=\cup Z_j,$ with $Z_j$ the singular set determined by $l^j$ as above, and the $j$th factor of $S^1$ acts on the $i$th variable as $x_i\rightarrow e^{\sqrt{-1}l_i^j\theta_j}x_i.$ For clarity, throughout the paper only one Kahler parameter will be considered. 

To get at the examples we want to study, it is necessary to construct the $B$ model geometry $\tilde X_a$ to $\tilde W_r.$ Local mirror symmetry techniques have been known for some time \cite{CKYZ}, and the current formulation differs slightly (e.g. \cite{AV}). We will see that both methods give identical results, in terms of what is considered here. This was also noticed by Orlov \cite{O}. 

We start by finding an intermediate space $\tilde X_{\psi}.$ Classical mirror symmetry postulates a map between the complexified Kahler parameters of $A$ model geometry and the complex moduli of $B$ model geometry, so in the same spirit, we must complexify the Kahler modulus $r$ of $\tilde W_r.$ This is simply $\psi = r+i\theta,$ where $\theta$ is the angle determined by the $S^1$ action. Then according to Vafa et al., if $x,z\in\C$ and $y_i\in\C^*$ for $i=0\dots3,$ the mirror of $\tilde W_r$ is a hypersurface 

$$ 
\tilde X_{\psi} = \{xz+\sum_{i=0}^3 y_i=0 | \prod_{i=0}^3 y_i^{l_i}=e^{-\psi}\}. 
$$ 

This has one more complex dimension than desired, so in computations we must specialize to a patch where $y_i=1$ for some $i$. 

The relationship between this and earlier local mirror symmetry is the factor of $xz$ out front, as well as the condition that $y_i\in\C^*$. This second modification changes the invariance of the period integrals, and thus does affect the form of the PF operators slightly. We are then led to write the $B$ model as a Riemann surface 

$$ 
\{(y_0, y_1, y_2)\in (\C^*)^3 | \sum_{i=0}^3 y_i=0 , \prod_{i=0}^3 y_i^{l_i}=e^{-\psi}, y_0=1\}. 
$$ 

As found in \cite{O}, the category of $B$ branes (i.e., holomorphic submanifolds) is unchanged with the addition of the $xz$ summand. 

It was mentioned earlier that the PF equations would be derived by enlarging the moduli space and then quotienting at the end; hence, we finally arrive at our $B$ model space for noncompact Calabi- Yaus, in hypersurface form: 

\begin{equation} 
\label{noncptB} 
\tilde X_a = \{xz + \sum_{i=0}^3 a_iy^{m_i} =0\}, 
\end{equation} 
where $a_i$ and $y_i$ are in $\C^*$ for all $i,$ and $x,z\in\C.$ Here it is understood that the $y^{m_i}$ part was gotten by solving the relation $\prod_{i=0}^3 y_i^{l_i}=1$ for one of the $y_i$ variables. The modulus $e^{-\psi}$ is now absorbed by the $a_i.$ 

With our space at last defined, we can see how the definition of the hypersurface $Y_{a,b}$ for compact examples was found. First, let's take a look at special Lagrangian submanifolds of (\ref{noncptA}); since the $B$ model hypersurfaces are mirror to these special Lagrangians, this will tell us which holomorphic submanifolds to look at, similarly to the way we found (\ref{noncptB}) from the $A$ model. 

A nice class of SLags on $\tilde W_r$ was exhibited in \cite{AV}, and these are: 

$$ 
\tilde L_{r,c} = (\tilde W_r \cap\{\sum_i q_i^1|x_i|^2=c, \sum_i q_i^2|x_i|^2=0\})/S^1, 
$$ 
where this time the $S^1$ acts as $x_i\rightarrow e^{\sqrt{-1}q^1_i\phi}x_i$ for each $i$. Again, we must insist that $\sum_{i=0}^3q_i^j=0$ for $j=1,2.$ 

The equivalence of the $A$ and $B$ model string theories dictates that there must be a holomorphic submanifold of $\tilde X_{\psi}$ which is mirror to $\tilde L_{r,c}.$ Note the strong formal similarity between the conditions 

$$\sum_i l_i|x_i|^2=r, \ \ \ \sum_i q_i^1|x_i|^2=c.$$ 

Then the mirror to $\tilde L_{r,c}$ is given, not surprisingly, as 

$$ 
\tilde Y_{\psi, c}' = \tilde X_{\psi} \cap \{\prod_{i=0}^3 y_i^{q_i^1}=e^{-c}, \prod_{i=0}^3 y_i^{q_i^2}=1\}. 
$$ 

This is exactly the mirror $B$ brane as described in \cite{AV}. 

To connect this physical fact with the space $\tilde Y_{a,b}$ to be given below, we must account for the observation of \cite{LM}, \cite{LMW} that it suffices instead to consider the hypersurface 

$$ 
\tilde Y_{\psi,c} = \tilde X_{\psi}\cap \{\prod_{i=0}^3 y_i^{q^1_i}=e^{-c}\}. 
$$ 

This assumption is not so outlandish, as the excised condition $\prod_{i=0}^3 y_i^{q_i^2}=1$ contains no moduli information. However, later on we will see that we can also consider the full mirror geometry, together with  $\prod_{i=0}^3 y_i^{q_i^2}=1$, and obtain the same results.
 
Next, in all cases practically arising in physics, we have that $q^1 = e_i-e_j,$ where $\{e_i\}$ is the standard basis in $\mathbb{R}^4.$ Thus, the hypersurface condition becomes 

$$ 
y_iy_j^{-1}=e^{-c} \ \ \longleftrightarrow \ \ y_i-e^{-c}y_j =0. 
$$ 
Hence, by enlarging the moduli spaces, we at last obtain our hypersurface: 

\begin{equation} 
\label{noncpthyper} 
\tilde Y_{a,b} = \tilde X_a \cap \{b_0y_0 + b_1y_1=0\}. 
\end{equation} 

The $y_i$ and $y_j$ have been simply put as $i=0,j=1$ for simplicity, but it is clear that for different choices of special Lagrangian on the $A$ side, (\ref{noncpthyper}) can be easily modified accordingly.

\section{ "Periods" on relative cohomology.}

\subsection {Relative Periods.}

This section reviews periods and the residue construction for the usual, non- relative setting. We follow \cite{M} and \cite{CK}; note that these constructions apply to the compact case. Let $\rho : \chi \longrightarrow C$ be a family of smooth projective algebraic varieties of dimension $n$, where $C = {\Delta}^* .$ Fix $z_0 \in C$ and set ${\rho}^{-1}(z_0) = X .$ Also, choose a basis $\{ {\gamma}_1 , ... , {\gamma}_r \}$ for $H_n (X),$ and a holomorphic $(n,0)$ form $\Omega$ on $X .$ The \itshape periods of X \upshape are then 

$$ \int_{{\gamma}_1} \Omega,..., \int_{{\gamma}_r} \Omega .$$ 

One can locally extend the form $\Omega$ to a family $\Omega (z)$ of forms on the fibers $X_z ,$ where $z$ is the coordinate on $C ,$ as well as the cycles, to get ${\gamma}_i (z) .$ Then $\Omega (z)$ will be a section of a bundle $ {\cal F}^n ,$ which is a subbundle of $R^n \rho_* \C \otimes {\cal O}_C $ with fiber $F_z^n (X) $ (this is the $n$th subspace of the usual Hodge filtration on $X$). Then the Picard- Fuchs equations arise by applying the canonical flat Gauss- Manin connection $\nabla$ defined on $R^n \rho_* \C \otimes {\cal O}_C $ to the period integrals: 

$$ \frac{\rm \it d}{\it dz} \int_{{\gamma}_i (z)} \Omega (z) = \int_{{\gamma}_i (z) } \nabla_{ d / dz } \Omega (z) . $$ 
This holds from the local constancy of the cycles on $X$. Relations among the periods can then be found as relations between the derivatives of $\Omega (z)$, modulo exact forms. 

Next, we want to do the corresponding thing for the present case. Let $\pi : \chi \longrightarrow S$ be as in section $2 ,$ together with the decreasing Hodge filtration of holomorphic vector bundles $ {\cal F}^{\bullet} $ over $S . $ Say $\Omega (z)$ is a section of ${\cal F}^n$; then $\Omega (z) \in H^n(X,Y)_z$ for each $z \in S .$ Then choosing a basis $\{ \Gamma_i \}_{i=0}^s $ for the homology $H_n(X,Y)_z$ (that is, $n$ cycles in $X$ that are disjoint from $Y$) for some fixed $z$, it is natural to consider, in the above context, the \itshape relative periods \upshape \cite{LMW}: 

$$\int_{{\Gamma}_1} \Omega (z),..., \int_{{\Gamma}_s} \Omega (z).$$ 

Exactly the same arguments as above apply, and again we use the splitting of the relative homology sequence to show that the $\Gamma_i (z)$ are locally constant in $z.$ Hence, we are looking for Picard Fuchs operators in the variables $z_1$ and $z_2$ which annihilate a certain element of $H^n(X,Y) .$

\subsection {Definition of the Relative Cohomology Class.}

For motivation, note that the right exactness $$ H^n(X,Y) \longrightarrow H^n(X) \longrightarrow 0 $$ means that any form in $\h$ can be realized as the pullback of a form in $H^n(X) .$ Thus, we will first review the construction of such forms on $X,$ and see how this can be used to produce a relative form of the desired type. 

So, return to the first case in the previous subsection, in which $\rho : \chi \longrightarrow C$ was a family of smooth projective varieties of dimension $n.$ For simplicity, suppose these $\rho^{-1} (z)$'s arise as hypersurfaces in $\p_{\Delta},$ and choose a definite $X = \rho^{-1}(z).$ Say that this $X$ is defined by a Laurent polynomial $f$, and let $\Omega_0$ be the canonical $(n+1,0)$ form on $\p_{\Delta}$, which is given by \cite{B}: 

$$ \Omega_0 = \frac{dy_0}{y_0} \wedge...\wedge \frac{dy_{n}}{y_{n}}.$$ 
Recall that the $\{y_i\}$ are coordinates on $T_N$.

It is well known that the residue map 

$$Res : H^{n+1,0}(\p_{\Delta}-X) \longrightarrow H^{n,0}(X)$$ 
(to be defined below) is surjective in this case, so that any holomorphic form on $X$ can be obtained from one on $\p_{\Delta}.$ With the data already given, we have $\Omega_0 / f \in H^{n+1,0}(\p_{\Delta}-X).$ Now, let $\gamma$ be an $n$ cycle in $X$ and let $T(\gamma)$ be a tube over $\gamma$, i.e., an $S^1$ bundle over $\gamma$ contained in the normal bundle of $X$ in $\p_{\Delta}.$ Then we can define the residue map by 

$$ \int_{\gamma} Res \sl ( \frac{ \Omega_0}{ f}) = \int_{T(\gamma)} \frac{\Omega_0}{f} .$$ 

As $T(\gamma)$ is independent of $z,$ we can freely differentiate under the integral sign of the right hand side of the above equation to find the Picard Fuchs equations, up to exact forms. Hence, it suffices to consider $\Omega_0 / f$ as a "form" on $X.$ 

Now we switch to the original relative setting, with $\pi : \chi \longrightarrow S$ as in previous sections, and fix a particular pair $(X, Y) = \pi^{-1}(z).$ Recall that here, the submanifold $Y$ is specified by $Y = X \cap \{g=0\}.$ With the construction at the beginning of this section, it is natural to proceed in the following way. First, in practice we will be computing with forms in $H^n_c(X-Y)$, and the compact support will be realized as a cutoff in the geometry integral. Then, for any $\Omega_X \in H^{n,0}(X)$, a form defined on $H^n(X-Y)$ is $\Omega_X / g .$ Combining this with the above gives the form that we'll use in the derivation of the Picard Fuchs equations: 

$$ \Omega = \frac {\Omega_0}{fg} . $$ 

An immediate observation is that any operator which annihilates the form $\Omega_0 / f$ will also annihilate $\Omega,$ because the polynomial $g$ does not depend on the moduli of $X$ (even though the space $Y$ does).

\subsection {Derivation of the GKZ operators on the enlarged moduli space.}

From this point forward, all computations will be carried out in the toric setup of section 2.4. Again, these derivations are for compact $X.$ So, with the conventions of that section, we consider the family of Calabi- Yau toric hypersurfaces $$X_a = \{ f = \sum_{i \ge 0} a_i y^{m_i} = 0\},$$ 
which are sections of the anticanonical bundle of $\p_{\Delta},$ and the hypersurface in $X_a$ 

$$ Y_{a,b} = X_a \cap \{ g = b_0 y_0 + b_1 y_1 = 0 \} .$$ One more assumption must be made on the form of $X_a$ here, which is that $m_0 = (1,0,...,0)$ and $m_1 = (0,1,0,...,0).$ As this is satisfied in every case arising from physical considerations, it puts no restriction on the class of examples on which mirror maps can be constructed. 

The moduli space is supposed to consist of $c = (a,b) = (a_0,...,a_n,b_0,b_1)$, which is in fact larger than the $2$ dimensional setup at the beginning of the paper; this will be remedied in an upcoming section. For now, we construct the GKZ operators directly on the moduli space $c.$ Using the form previously defined, 

$$ \Omega_0 = \frac{dy_0}{y_0} \wedge...\wedge \frac{dy_{n}}{y_{n}},$$ 
the standard construction is to write a form on $\p_{\Delta} - X$ as $\Omega_X = \Omega_0 / f .$ From this, one again has a lattice vector of relations 

$$\Lambda = \{l \in {\Z}^{n+1}: \sum_i l_i m_i = 0, \sum_i l_i = 0 \},$$ 
as was used to define the canonical coordinates in section 2.4. For our situation, $\Lambda$ will always be a $1$ dimensional space. For a suitable choice of basis vector (as described by the Batyrev mirror) $l^1$ of $\Lambda,$ one can define a GKZ operator 

$${\cal L}_1 = \prod_{{l_k^1}>0} {\partial}_{a_k}^{l_k^1} - \prod_{{l_k^1}<0} {\partial}_{a_k}^{-{l_k^1}},$$ 
which can be easily shown to annihilate the form $\Omega_X.$ 

Now, following the procedure outlined earlier, the form we want to define in order to derive an extended set of GKZ operators is given by 

$$\Omega = \frac {\Omega_X}{g} = \frac{\Omega_0}{(\sum_{i \ge 0} a_i y^{m_i})( b_0 y_0 + b_1 y_1) }. $$ 

This makes it clear that the operator ${\cal L}_1$ gotten from $\Omega_X$ also annihilates $\Omega,$ as $g$ is independent of the $a_i.$ Since we want new operators with $\Omega$ in their kernel, it is convenient to define a new set $\widehat{\Lambda},$ which consists of all $\widehat{l}^i \in \Z^{n+3}$ such that 

$${\cal L}_i \Omega = (\prod_{{\hat{l}_k^i}>0} {\partial}_{c_k}^{\hat{l}_k^i} - \prod_{{\hat{l}_k^i}<0} {\partial}_{c_k}^{-{\hat{l}_k^i}}) \Omega = 0.$$ 

The fact that the original operator ${\cal L}_1$ sends $\Omega$ to $0$ is captured by the inclusion map $\Z^{n+1} \hookrightarrow \Z^{n+3}$ applied to the set $\Lambda.$ Henceforth, for simpler notation we will drop the hats and just consider $\Lambda \subset \Z^{n+3}.$ 

From the defining equation of $\Omega$, one more operator is easily seen to annihilate it, namely 

$${\cal L}_2 = \partial_{b_0}\partial_{a_1} - \partial_{a_0}\partial_{b_1} .$$ This corresponds to a new element of $\Lambda,$ given by $(-1,1,0,...,0,1,-1).$ As will be shown shortly, the set $\{{\cal L}_1, {\cal L}_2\}$ is sufficient to determine the open string mirror map.

\subsection{Noncompact CY's.}

For a non- compact $\tilde{X_a},$ we are led instead to produce an $(n+2, 0)$ form on relative cohomology. This goes as follows. The $(n+3, 0)$ form on $\C^2 \times \p_{\Delta},$ which is used for the residue construction for relative cohomology, is $\tilde{\Omega_0} = dxdz \Omega_0.$ Then, using the same arguments as before, the form needed for derivation of the GKZ operators is

$$ 
\tilde{\Omega} = \frac{dx dz \Omega_0}{(xz + \sum_{i \ge 0} a_i y^{m_i})(b_0y_0 + b_1y_1)}. 
$$ 

This makes it clear that the same operators $\{{\cal L}_1, {\cal L}_2\}$ still send $\tilde{\Omega}$ to 0. Thus, the equations themselves are unchanged between the two cases at the level of GKZ operators. The Picard- Fuchs operators are, however, different, as will be described below.

\subsection{Another way of looking at relative cohomology.}

Recently, Hosono \cite{H} has given a description of periods on noncompact Calabi-Yaus, which can be used in the present setting to argue the naturality of the cohomology class $\tilde \Omega$ above. 

Start with the $B$ model hypersurface description given above: 

$$ 
\tilde X_{\psi} = \{xz+\sum_{i=0}^3 y_i=0 | \prod_{i=0}^3 y_i^{l_i}=e^{-\psi}\} 
$$ 
$$ 
=\{xz+f(z_1)=0 \}, 
$$ 
where we have set $z_1 = e^{-\psi}$ and used the condition $\prod_{i=0}^3 y_i^{l_i}=z_1$ to eliminate one of the $y_i$ variables in $xz+\sum_{i=0}^3 y_i=0$. 

Then, according to Hosono, we should define the periods of $\tilde X_{z_1}$ by the integrals 

$$ 
F(z_1)= \int_{\Gamma}\frac{dx dz \prod_i dy_i/y_i }{xz+f(z_1)} 
$$ 
where $\Gamma \in H_{n+1}(\C^2 \times (\C^*)^{n-1} - \{xz+f(z_1)=0\},\Z).$ The PF operators derived from the $F(z_1)$ are of course identical to those for $\Omega_0/f$ given above. 

Next, we would like to take the mirror geometry of the special Lagrangians found above seriously; namely, let 

$$ 
\tilde Y_{z_1, z_2}' = \tilde X_{z_1} \cap \{\prod_{i=0}^3 y_i^{q_i^1}=z_2, \prod_{i=0}^3 y_i^{q_i^2}=1\} 
$$ 
$$ 
= \tilde X_{z_1} \cap \{g(z_2)=0, h=0\} 
$$ 
$$ 
= \{(x,z,y_1,y_2)\in \C^2 \times (\C^*)^2|xz+f(z_1)=g(z_2)=h=0\}; 
$$ 
notice that $h= \prod_{i=0}^3 y_i^{q_i^2}-1$ has no moduli. Then, the corresponding thing to Hosono's work is to define the "relative periods" for noncompact Calabi-Yau manifolds by integrals 

$$ 
F(z_1,z_2) = \int_{\Gamma}\frac{dx dz \prod_i dy_i/y_i }{(xz+f(z_1))g(z_2)h}. 
$$ 
where here $\Gamma \in H_{n+1}(\C^2 \times (\C^*)^{n-1} - \{xz+f(z_1)=g(z_2)=h=0\},\Z)$ 

This formulation of relative cohomology eliminates the need to resort to physical arguments for why we ought to work with a hypersurface, and shows that the PF operators can be directly derived from the "honest" mirror of the special Lagrangian.

\section {Picard Fuchs equations.}

\subsection {Quotients of the enlarged moduli space.}

The next step is to transform the GKZ system into Picard Fuchs equations on the quotient space. First, though, we need a better understanding of the quotient spaces of $X_a$ and $Y_{a,b}$ on which the PF system lives. 

So, we first consider the natural actions on these spaces that we want to quotient out. The space $X_a = \{ f= \sum_{i \ge 0} a_i y^{m_i}= 0\}$ has a torus action $T$ and a $\C^*$ action defined by 

$$ T: y_i \longrightarrow \nu_i y_i$$ $$\C^* : f \longrightarrow c_1f.$$ 

As mentioned earlier, the set $\Lambda \subset \Z^{n+1}$ of relations among the $m_i$ defines a coordinate $ z_1 = \Pi_k{a_k}^{l_k^1}$; this is evident, because $z_1$ is invariant under $T$ and $\C^*.$ Thus, we can set 

$$X_{z_1} = X_a / (T \times \C^*).$$ 

Now, $Y_{a,b} = X_a \cap \{g= b_0y_0 + b_1y_1 = 0\}$ has a $T'$ action given by restricting $T$ to $Y,$ as well as an independent $\C^*$ action $g \longrightarrow c_2g.$ By instead considering $\Lambda \subset \Z^{n+3},$ and also including (as above) the vector $l^2=(-1,1,0,...,0,1,-1)$ into this set, we can define the coordinates 

$$z_i = \prod_{k=0}^n {a_k}^{l_k^i} {b_0}^{{l}_{n+1}^i} {b_1}^{{l}_{n+2}^i} = \prod_{k=0}^{n+2} c_k^{l_k^i},$$ 

For the vector $l^1 \in \Lambda$, the coordinate is the same as that on $X$. There is also the new coordinate $z_2$ defined by $l^2,$ which is the \itshape open string modulus. \upshape Then $(z_1,z_2)$ are invariant under $T, T'$ and both $\C^*$ actions, and therefore define coordinates on the total moduli space. We thus have 

$$Y_{z_1,z_2} = (X_a/(T \times \C^*)) \cap (\{g=0\}/(T' \times \C^*)).$$ 

Note that the form $\Omega$ from section 4 is not invariant under the two $\C^*$ actions. Hence, to define a form that descends to the quotient, simply set $\check{\Omega} = a_0b_1\Omega.$ We will drop the check in the sequel.

\subsection {(Problems with) the compact case.}

All of the above was carried out with the assumption that the techniques of \cite{LMW}, which were developed in the noncompact case, could be generalized in the natural way to include compact Calabi-Yaus. Here we will see that some nontrivial modification of \cite{LMW} is needed to accomodate compact examples.

With the quotients of $X$ and $Y$ out of the way, and a form $\Omega$ defined on $H^n(X-Y)_z,$ we can now see what the form of the Picard- Fuchs equations coming from the GKZ operators is. Compact solutions will be written first, with the noncompact modifications described afterwards. While the computations are straightforward, the physical interpretation of the results seems to be inconsistent. For the moment, we will carry out the program and comment on its difficulties afterwards. 

Note that, in order to have operators which annihilate the invariant $\Omega,$ we are obliged to modify the original ${\cal L}_1, {\cal L}_2$: 

$$ {\cal D}_1 = {\cal L}_1 {a_0}^{-1}{b_1}^{-1}, \ \ \ {\cal D}_2 = {\cal L}_2a_0^{-1} b_1^{-1}.$$ 

The Picard- Fuchs operators corresponding to these have already been worked out in \cite{HKTY}, so here we can directly write down their form. The convention used below is that $\theta_i = z_i\partial/\partial z_i$, with $i=1,2$. We have 

\begin{equation} 
\label{PF} 
{\cal D}_i = \prod_{ l_k^i > 0} \prod_{n=0}^{l^i_k - 1} (\sum_{i=1}^2 l_k^i \theta_i - n) - z_i\prod_{ l_k^i < 0}\prod_{n=1}^{-l_k^i} (\sum_{i=1}^2 l_k^i\theta_i - n). 
\end{equation} 
The product $ \prod_{ l_k^i > 0}$ is taken over $k.$

Now, we would like to find the general solutions of the relevant Picard- Fuchs equations, in the compact setting. For this case, we can compute them by direct evaluation of the period integrals, as follows. Let $\gamma$ be a contour enclosing $fg=0$ such that $|y_i| =1$ on $\gamma$ for all $i.$ Then the period 

$$\frac{1}{(2\pi i)^{n+1}} \int_{\gamma} \frac{a_0 b_1}{(\sum_{i \ge 0} a_i y^{m_i})(b_0y_0 + b_1y_1)} \prod_{i=0}^{n} \frac{dy_i}{y_i}$$ 
can be calculated by expanding $\Omega$ as a function of $(a_0^{-1},b_1^{-1})$ around the point $(0,0)$ and using the residue theorem. For applications, this is a valid approximation, which can be understood by the following. Since $\sum_k l_k^1 = 0,$ we can take $ l_0^1 < 0$. Also, $l^2_{n+2} < 0,$ and considering both of these we see that the limits 

$$ |a_0| \rightarrow \infty, \ \ |b_1| \rightarrow \infty $$ 
correspond to the maximally unipotent boundary point $z_1 = z_2 = 0$, by the formula for the coordinates $z_1, z_2.$

Going back to the expansion, if $c = (a_0,...,a_n,b_0,b_1)$, the answer we get from this is 

$$y_0(z) = \sum_{n_i} \frac{\prod_{ l_k^i < 0} (-\sum_{i=1,2}l_k^i n_i)!}{\prod_{ l_k^i > 0}(\sum_{i=1,2}l_k^i n_i)!} \prod_{k = 0}^n c_k^{\sum_i l_k^i n_i}$$ 

$$= \sum_{n_i} \frac{\prod_{ l_k^i < 0}(-\sum_{i=1,2}l_k^i n_i)!}{\prod_{ l_k^i > 0}(\sum_{i=1,2}l_k^i n_i)!}((-1)^{\sum_{ l_k^1 < 0}l_k^1} z_1)^{n_1} ((-1)^{\sum_{ l_k^2 < 0}l_k^2}z_2)^{n_2},$$ 
where the second line is gotten from the definition of the coordinates $z_1, z_2.$ Above, the sum is over all $n_i$ such that $\sum_i l_k^i n_i \ge 0 $. This solution can be further simplified by the observations that $l_{n+1}^1 = l_{n+2}^1 = 0$ and $l_{n+1}^2 = -l_{n+2}^2$. Then we have 

\begin{equation} 
\label{eqno3} 
y_0(z) = \sum_{n_i} \frac{(-\sum_{i=1,2} l_0^i n_i)!}{\prod_{k \in \{1,...,n\}} (\sum_{i=1,2} l_k^i n_i)!} ((-1)^{l_0^1 } z_1)^{n_1} ((-1)^{l_0^2 }z_2)^{n_2} 
\end{equation} 

By conjecture, this is the unique solution of the PF equations which is holomorphic at every maximally unipotent boundary point. This corresponds to the condition dim$M_0 = 1$ in the definition of maximally unipotent monodromy. Notice that this formula passes a consistency check from the closed string case: if $l^2$ is the $0$ vector, then the above expression agrees with equation (6.50) of \cite{CK}.

With this information, there is only one more component required to identify the open string mirror map, and that is the application of the Frobenius method to the solution $y_0$. So, consider the function 

$$ y_0 (z,\rho) = \sum_{n_i} c(n, \rho) ((-1)^{l_0^1} z_1)^{n_1 + \rho_1} ((-1)^{l_0^2 }z_2)^{n_2 + \rho_2}, $$ 
where 

$$ c(n, \rho) = \frac{\Gamma(-\sum_{i=1,2}l_0^i (n_i+ \rho_i) +1)}{\prod_{k \in \{1,...,n\}} \Gamma(\sum_{i=1,2}l_k^i (n_i + \rho_i) +1)}.$$ 
From this, it is possible to construct two more PF solutions: 

$$y_i = ( \frac{\partial}{\partial \rho_i} y_0(z,\rho))|_{\rho = 0} .$$ 
These can be shown to satisfy $$y_k = y_0 log ((-1)^{l_0^k } z_k) + \widetilde{y_k},$$ simply by applying the identity

$$\partial_{\rho_k} ((-1)^{l_0^k} z_k)^{n_k + \rho_k} = log((-1)^{l_0^k} z_k) ((-1)^{l_0^k } z_k)^{n_k + \rho_k}.$$ 
It then follows immediately that 

\begin{equation} 
\label{eqno4} 
\partial_{\rho_k} y_0(z,\rho) = y_0(z,\rho)log((-1)^{l_0^k} z_k) + 
\sum_{n_i} (\partial_{\rho_k} c(n, \rho))((-1)^{l_0^1 } z_1)^{n_1 + \rho_1}((-1)^{l_0^2 }z_2)^{n_2 + \rho_2} 
\end{equation} 
which is as claimed.

In direct analogy with the closed string case, we would be led to define the flat coordinates $t_1, t_2$ on the $(X,Y)$ moduli space for a compact Calabi- Yau as follows: 

$$t_i(z) = \frac{y_i(z)}{y_0(z)}, \ \ i = 1, 2 \ ;$$ then $z_i \longrightarrow t_i(z)$ would be called the open string mirror map. 

The problem with all this, of course, is that, to date, there are no examples of compact open string mirror symmetry. We have simply proceeded analogously to the way one would when adapting noncompact closed string mirror symmetry techniques to the compact closed string case. As we will see below, such a naive extension is in fact not possible.

\subsection{Noncompact Solutions.} 

From the explicit form of $\tilde{X_a},$ we see that the same GKZ differential equations of the last section hold for the relative cohomology of $\tilde{X_a}.$ However, as in the closed string case, we find slight differences in the form of the Picard- Fuchs equations, and hence of the solutions. Following \cite{LM}, the noncompact Picard- Fuchs system is 

\begin{equation} 
\label{noncptPF} 
{\tilde{\cal D}}_i = \prod_{ l_k^i > 0} \prod_{n=0}^{l^i_k - 1} (\sum_{i=1}^2 l_k^i \theta_i - n) - z_i \prod_{ l_k^i < 0}\prod_{n=0}^{-l_k^i-1} (\sum_{i=1}^2 l_k^i\theta_i - n). 
\end{equation} 

These operators differ from those of (\ref{PF}) because in the noncompact case, we use the GKZ operators ${\cal{L}}_i$ rather than ${\cal{L}}_i a_0^{-1} b_1^{-1}$ to derive Picard- Fuchs equations. This is on account of the invariance of ${\cal{L}}_i$ in the noncompact case, as discussed in \cite{CKYZ}.

The solutions to this can be easily found simply by shifting all gamma functions appearing in the numerator of the compact periods by $-1$. Notice that there is always a constant solution 1 to the above system. Then, again through the Frobenius method, all solutions may be derived from

$$ 
\tilde{y_0}(z,\rho)=1 + \rho_1 l_0^1 \sum_{n_1 \ge 1, n_2 = 0} \frac{(-1)^{l^1_0 n_1}(l_0^1n_1 -1)!}{\prod_{k \in \{1 \dots n \}} (l^1_k n_1)!} z_1^{n_1} + 
$$ 

\begin{equation} 
\label{noncpt} 
\sum_{n_1 \ge 0, n_2 \ge 1} \frac{(-1)^{l_0^1n_1 } \Gamma(-\sum_i l_0^i(n_i+\rho_i))\Gamma(-\sum_i l^i_{n+1}(n_i + \rho_i))}{\prod_{k \in \{{1 \dots n, n+1}\}}\Gamma(\sum_i l_k^i(n_i + \rho_i) + 1) }z_1^{n_1+\rho_1}z_2^{n_2+\rho_2} 
\end{equation} 

To understand this formula, recall from the compact case that after normalization, we were left with a constant solution $1 = y_0/y_0.$ From the form of $\tilde{y_0},$ the leading 1 represents the constant solution, and the next term is the correction to the closed string mirror map. This correction is generally found as the logarithmic solution of the PF eqns modulo the log term. Now, from the theory of open strings \cite{AV} \cite{LM}, it is conjectured that the moduli space coordinates $\{z_i\}$ receive only closed string corrections. With the above formula in mind, this means that only terms with $n_2 = 0$ can correct the flat coordinates, and thus both logarithmic open string PF solutions must be of the form 

$$ 
t_i(z) = log(z_i) + k_i l_0^1 \sum_{n_1 \ge 1, n_2 = 0} \frac{(-1)^{l^1_0 n_1}(l_0^1n_1 -1)!}{\prod_{k \in \{1 \dots n \}} (l^1_k n_1)!} z_1^{n_1} 
$$ 
for $i=1,2$ and $k_i \in \mathbb{R}.$ The coefficients $k_i$ are determined simply by the requirement that the $t_i$ satisfy the extended PF system. The exact form of the $t_i$ can be found in $\cite{LM}.$ 

As in the noncompact case for closed strings, here we can write down the mirror map directly, without having to resort to a ratio of power series. The reason is that since the system (\ref{noncpt}) has the constant solution 1, we can think of the period vector as normalized from the outset. 

The meaning of the above PF system has been recently clarified \cite{H}. Namely, for the relative cohomology class $\Omega_0 /fg,$ the noncompact PF operators annihilate the integrals 

$$ 
\int_{\Gamma}\frac{\Omega_0}{fg}, 
$$ 
where $\Gamma \in H_{n+1}(\p_{\Delta}-\{fg=0\},\Z).$ This explains why it is necessary to include the condition $\Omega_0 /fg \in H^n_c(X-Y),$ and therefore why $\h$ is the correct group to look at in the first place. 

\subsection{Superpotentials and Open Gromov Witten invariants.}

Finally, we would like to compute the superpotential $W(z_1,z_2)$, analogously to the computation of the prepotential ${\cal F}$ for the closed string case. The purpose of this is the same as that of the prepotential computation in \cite{M}: the expansion of the superpotential in terms of the flat coordinates $t_i$ defined above is conjectured to have integral coefficients. These integers will then give predictions for the number of open Gromov Witten invariants, whose definition is still underway \cite{KL}. 

Stated more precisely, the superpotential is conjectured to be of the form \cite{LMW} \cite{AV} 

\begin{equation} 
\label{eqno45} 
W(t_1,t_2) = \sum_{n_1,n_2} N_{n_1,n_2} \sum_k \frac{1}{k^2} \prod_{j=1,2} e^{2 \pi i k n_j t_j}. 
\end{equation} 
Above, the coefficients $N_{n_1, n_2}$ are supposed to be integers giving the number of open Gromov- Witten invariants of certain topologies. 

Before any such enumerative test can be carried out, we need an algorithm to produce the superpotential. First, the compact case. Then following the Frobenius method, the answer is easy: 

\begin{equation} 
\label{eqno5} 
W(z_1, z_2) = \frac{1}{y_0}\sum_{n_1,n_2} \frac{\partial}{\partial \rho_1} \frac{\partial}{\partial \rho_2} c(n, \rho) |_{\rho = 0} ((-1)^{l_0^1}z_1)^{n_1}((-1)^{l_0^2} z_2)^{n_2}. 
\end{equation} 
To see how this comes about, refer back to (\ref{eqno4}). Then 

$$ \frac{\partial}{\partial \rho_1} \frac{\partial}{\partial \rho_2} y_0(z, \rho) = log((-1)^{l_0^1}z_1)log((-1)^{l_0^2} z_2) y_0(z, \rho) 
$$ 
$$ 
+log((-1)^{l_0^1}z_1)\sum \partial_{\rho_2} c(n, \rho)((-1)^{l_0^1}z_1)^{n_1 + \rho_1}((-1)^{l_0^2}z_2)^{n_2 + \rho_2} 
$$ 
$$+log((-1)^{l_0^1}z_2)\sum \partial_{\rho_1} c(n, \rho)((-1)^{l_0^1}z_1)^{n_1 + \rho_1}((-1)^{l_0^2}z_2)^{n_2 + \rho_2} 
$$ 
$$ 
+ \sum \partial_{\rho_1} \partial_{\rho_2} c(n, \rho) ((-1)^{l_0^1}z_1)^{n_1 + \rho_1}((-1)^{l_0^2}z_2)^{n_2 + \rho_2} 
$$ 

According to \cite{LMW}, this will give the superpotential modulo the logarithmic terms after dividing by $y_0,$ since the log terms are supposed to correspond to the classical superpotential, which is $0$ here. After dropping the logarithmic terms, dividing by $y_0$ and setting $\rho = 0$, we get (\ref{eqno5}). 

To better understand this definition, we can consider the relative period vector 

$$ 
\Pi (z) = (y_0, y_1, y_2, y_0 W). 
$$ 
It is conventional \cite{HKTY} to normalize this vector by dividing through by $y_0,$ which gives 
$$ 
\Pi(z) =y_0 (1, t_1, t_2, W). 
$$ 
The middle two entries would be defined by the yet unknown compact mirror map. 

So, the above gives the framework that should hold if the techniques of \cite{LMW} can be adapted to the compact case. However, the family of submanifolds is holomorphic, and physically it is expected that the superpotential will be $0$ in this case. Thus, the interpretation of the compact calculation is unclear, though it is possible that it can be thought of in the context of closed string mirror symmetry on Calabi-Yau $4$-folds. At any rate, the natural generalization of \cite{LMW} to compact Calabi-Yaus seems to be inconsistent, as demonstrated by the above superpotential computation.

Now, we move on to the noncompact case. Here, a much more explicit formula for the superpotential can be given, which eases comparison with earlier computations, e.g. \cite{GZ}. First refer back to equation (\ref{noncpt}). The leading 1 and the middle term together give the constant and logarithmic solutions respectively, as argued last section. In order to find the other solutions, then, we need only Taylor expand the third term in $\rho_1, \rho_2$ about $(0,0).$ The first term of this expansion will be the double logarithmic solution, which has the form 

\begin{equation} 
\label{doublelog} 
\tilde{W}(z_1, z_2) = \sum_{n_1 \ge 0, n_2 \ge 1} \frac{(-1)^{l_0^1n_1}(\sum_il_0^in_i -1)!}{l^2_{n+1}n_2 \prod_{k \in \{1\dots n\}}(\sum_il^i_kn_i)! } z_1^{n_1}z_2^{n_2}. 
\end{equation} 
Note that we are using a different sign convention from the compact case, so that $(-1)^{l_0^1 n_1 + l_0^2 n_2}$ is not included. 

For a consistency check, we can take the limit in which $n_1 = 0.$ This is expected to correspond to the case of $\C^3$. What we get from this is 

$$ 
\sum_{n_2 \ge 1} \frac{1}{n_2^2} z_2^{n_2}, 
$$ 
which is in fact the superpotential for the mirror of $\C^3$ \cite{AKV}.

\subsection{Closed string superpotentials.}

It is also known that superpotentials can be generated in the context of closed strings \cite{V}. For usual closed string theory, we have $N=2$ supersymmetry, which in the present setting (i.e., a 1 complex dimensional moduli space) means that the period integral would take the form 

\begin{equation} 
\label{periodvector} 
\Pi_0 = (1, t, d{\cal F} /dt,\dots) 
\end{equation} 
after normalization by the holomorphic solution $y_0$. Here $\cal F$ is the prepotential of the closed string theory. More generally, if our complex moduli space is $k$ dimensional, we would find a period integral 

$$ 
\Pi_0 = (1, t_1,\dots,t_k,\partial {\cal F} / \partial t_1,\dots,\partial {\cal F} / \partial t_k,\dots), 
$$ 
and $N=2$ supersymmetry would then mean that the functions $\partial {\cal F} / \partial t_i$ could be integrated to a single prepotential $\cal F$.

If we are using the open string moduli space (i.e, the complex moduli space together with the moduli of a holomorphic submanifold), then the $N=2$ supersymmetry is broken to $N=1,$ and we are forced to use the lesser information provided by the superpotentials defined in the preceding section. 

There is another way to break supersymmetry from $N=2$ to $N=1$, and that is through the inclusion of flux through the relevant holomorphic submanifold. In this case, the string theory has a superpotential $W = d{\cal F} /dt;$ here, there is no open string modulus, so the parameter $t$ unambiguously defines the flat coordinate on the complex moduli space. In fact, from the form of (\ref{periodvector}), it is evident that $W$ and $d {\cal F} / dt$ are both the double logarithmic solutions of the closed string Picard- Fuchs system, and hence are naturally identified. Thus, the same techniques presented above compute closed string superpotentials. 

For the simplest example of this, consider the total space ${\cal O} (-1) \oplus {\cal O} (-1) \rightarrow \p^1. $ This can be torically defined by the single vector $l = (1,1,-1,-1),$ and the associated noncompact Picard- Fuchs system has a double logarithmic solution given by 

$$ 
W(z_1) = \sum_{n\ge 1} \frac{1}{{n_1}^2} z_1^{n_1}. 
$$ 
This is well known to be the correct result for this space. Note, however, that while this solution can be obtained by naively applying the Frobenius method to the function

$$
\omega_0(z,\rho)=\sum_{n\ge 0}\frac{1}{\Gamma(1+n+\rho)^2\Gamma(1-n-\rho)^2} z^{n+\rho},
$$
the above $W(z_1)$ is not a solution of the PF system associated to the mirror of the local $\p^1$ geometry. This inconsistency will be addressed in a future paper \cite{BF}.

\section{Example : $K_{\p^2}.$}

In order to show that the above theory produces what it's supposed to, we will here outline the computation for the case of the mirror of $K_{\p^2}$. The following two subsections will specifically address consistency with the papers \cite{LM} and \cite{GZ}, respectively.

\subsection{Picard- Fuchs operators} 

Let $[y_0,y_1,y_2]$ be homogeneous coordinates on $\p^2,$ and consider a coordinate patch where $y_0 = 1.$ Let $a_0,...,a_3 \in \C^*,$ and $x,z \in \C.$ We have 

$$\tilde{X_a} = \{\tilde{f}(a) = xz+ f(a) = xz+ a_0 + a_1y_1 + a_2y_2 + a_3y_1^{-1}y_2^{-1} =0\}$$ and 

$$Y_{a,b} = \tilde{X_a} \cap \{g= b_0 + b_1 y_1= y_2-1= 0\}.$$

For fixed $a,$ $f(a) = 0$ defines a section of ${\cal O}(3) \longrightarrow \p^2$. Note that the points $\{m_i\} = \{(0,0), (1,0), (0,1),(-1,-1)\}$ given by the exponents of the $y_i$ in $f$ are the vertices of the polyhedron for $K_{\p^2}.$ Now, the lattice of relations $\Lambda \subset \Z^4$ among the $m_i$ is given by the single vector $(-3,1,1,1),$ which is then interpreted as a vector in $\Z^6$ by $l^1=(-3,1,1,1,0,0).$ The extended set $\Lambda \subset \Z^6$ consists of $l^1$ and $l^2 = (-1,1,0,0,1,-1)$ ,as derived earlier. Since the moduli coordinates are $(a_0,a_1,a_2,a_3,a_3,b_0,b_1),$ $\Lambda$ defines variables 

$$ z_1 = \frac{a_1a_2a_3}{a_0^3}, \ \ \ \ z_2 = \frac{a_1b_0}{a_0b_1} $$ 
on the reduced moduli space of the quotients of $\tilde{X_a}$ and $Y_{a,b}.$ We also have that the vectors $l^1$ and $l^2$ produce GKZ operators 

$${\cal L}_1 = \partial_{a_1}\partial_{a_2}\partial_{a_3} - \partial_{a_0}^3,$$ 

$${\cal L}_2 = \partial_{a_1}\partial_{b_0} - \partial_{a_0}\partial_{b_1} .$$ 
The invariant form we are using is $\Omega = \Omega_0 /\tilde{f}g,$ where 

$$\Omega_0 = dx\wedge dz \wedge \frac{dy_1}{y_1}\wedge \frac{dy_2}{y_2}.$$ 

Hence the operators $ {\cal D}_1 ({\cal D}_2) = {\cal L}_1 ({\cal L}_2 )$ will yield the Picard- Fuchs equations. 

Then, the operators ${\cal L}_1, {\cal L}_2$ annihilate the integrals 

$$ 
\int_{\Gamma} \frac{dx\wedge dz \wedge \frac{dy_1}{y_1}\wedge \frac{dy_2}{y_2}}{ (xz+ a_0 + a_1y_1 + a_2y_2 + a_3y_1^{-1}y_2^{-1})(b_0 + b_1 y_1 )(y_2 - 1)} 
$$ 
with $\Gamma \in H_4(\C^2 \times (\C^*)^2 - \{fg=0\}, \Z).$ 

From the equations defining the moduli variables $z_1$ and $ z_2$, if we let $\theta_i = z_i \partial/ \partial z_i$, we have the identities 

$$\delta_{a_2}= \delta_{a_3} = \theta_1, \ \ \ \ \delta_{a_1}= \theta_1 + \theta_2, \ \ \ \ \delta_{a_0} = -3\theta_1 - \theta_2, \ \ \ \ \delta_{b_0} = -\delta_{b_1} = \theta_2.$$ 
The derivations of the noncompact operators, ala \cite{LM}, lead one to find

$${\cal D}_1 = (\theta_1)^2(\theta_1 + \theta_2) + z_1(3\theta_1 +\theta_2)(3\theta_1 +\theta_2 + 1)(3\theta_1 +\theta_2 +2).$$ 
as well as 

$${\cal D}_2 = (\theta_1 + \theta_2)\theta_2 - z_2(3\theta_1 + \theta_2)\theta_2 .$$ 

Note that these are the same operators derived in \cite{LM}, \cite{LMW}(up to a minus sign in the $z_2$ variable), though with different techniques. In particular, this method provides a more mathematical explanation for the appearance of the $\theta_2$ factor on the right of ${\cal D}_2$. To produce exactly what was found in \cite{LM}, we need only use instead the vectors $l^1, -l^2$ as a basis for $\Lambda.$ Hence we find agreement with previous calculations of this example. The explanation of the minus sign on $l^2$ is that on the $A$ model, the Lagrangian submanifold ends on a different leg of the toric graph than was chosen for the calculations of section 5. The flopped phase $l^2 \rightarrow -l^2$ thus reflects different data on the mirror. 

The paper \cite{LM} contains the calculation of the superpotential (which is different from the one outlined above) and tables listing predicted open Gromov- Witten numbers. 

\subsection{Superpotential Computation.} 

Next, let's look at how the double logarithimic solution for the noncompact Picard- Fuchs system of the mirror of $K_{\p^2}$ compares with the results of \cite{GZ}. Writing the superpotential down for this theory is simply a matter of inserting $l^1, l^2$ into formula (\ref{doublelog}). The result of this is 

\begin{equation} 
\label{goat} 
W_{K_{\p^2}} = \sum_{n_1 \ge 0, n_2 \ge 1}\frac{(-1)^{n_1}(3n_1+n_2-1)!}{n_2(n_1+n_2)!(n_1!)^2}z_1^{n_1}z_2^{n_2}. 
\end{equation} 
Notice that this is a solution of the system $\{{\cal D}_1, {\cal D}_2\} $ found in the last section. 

To see explicitly the agreement with \cite{GZ}, we need to recall the notation of that paper. From page 14, we have the equation 

\begin{equation} 
\label{poop} 
\partial_{\hat u}^2 W(Q,y) = \sum_{w\ne0, d\ge0}y^w r^{-w}q^d(-1)^dw\frac{(3d+w-1)!}{(d!)^2(w+d)!}, 
\end{equation} 
where $y=e^{\hat u},q= Q/r^3 =e^{-t}.$ Further, $r= e^{\Delta/3}$ where $\Delta$ is a solution of the Picard-Fuchs equation 

$$ 
3\theta^3 + 3z\theta(3\theta +1)(3\theta+2). 
$$ 
In order to connect with the notation of section 6.1 above, we have 

$$z_1 = e^{-t}, z_2 = e^u$$ 
with the relation between $u$ and $\hat u$ given as $\hat u - \Delta/3 = u.$ This last equation disagrees with \cite{GZ} by a summand of $i\pi,$ but as pointed out in \cite{AV}, this is not fixed by mirror symmetry. The choice made here reflects consistency with that of \cite{LMW}. 

With all these notations out of the way, we integrate (\ref{poop}) twice in $\hat u$ to find the superpotential, which looks like 

$$ 
W(q,y) = \sum_{w\ne0, d\ge0}(y/r)^w q^d(-1)^d \frac{(3d+w-1)!}{w(d!)^2(w+d)!} 
$$ 
This is exactly the superpotential that was found in (\ref{goat}) above.


\end{document}